\documentclass[aps,pra,twocolumn,showpacs,notitlepage]{revtex4-1}

\usepackage{graphicx}
\begin{document}

\title{The quantum laws of physics:\\ a new description of dynamics and causality}

\author{Holger F. Hofmann}
\email{hofmann@hiroshima-u.ac.jp}
\affiliation{
Graduate School of Advanced Sciences of Matter, Hiroshima University,
Kagamiyama 1-3-1, Higashi Hiroshima 739-8530, Japan}
\affiliation{JST, CREST, Sanbancho 5, Chiyoda-ku, Tokyo 102-0075, Japan
}

\begin{abstract}
It is possible to completely explain all aspects of quantum mechanics by expressing the relations between physical properties in terms of complex conditional probabilities (Phys. Rev. A 89, 042115(2014)). These fully deterministic probabilities replace the classical notions of phase space geometries and continuous trajectories with a more accurate description that takes into account the role of dynamics in the emergence of reality. We can then understand why so many previous attempts to find a detailed explanation of quantum phenomena have failed: the assumption of a static reality breaks down when the interaction needed to obtain a real effect exceeds Planck`s constant. Beyond that limit, complex conditional probabilities are the only valid expression of the fundamental laws of physics. 
\end{abstract}


\maketitle

\section{Introduction}

The problem with quantum mechanics is that it fails to explain the dynamics of physical objects in terms of changes in the observable properties. The implicit claim is that the motion of an individual object is fundamentally unobservable, suggesting that trajectories might exist as ``hidden'' realities without any observable effects. However, the formalism of quantum mechanics does make very specific predictions about the relation between different physical properties even if these properties cannot be measured at the same time. Unfortunately, the mathematical formalism that provides these predictions is not based on empirical concepts, but originated from a rather daring identification of stationary states with orthogonal directions in Hilbert space. Importantly, this ad hoc definition of states cannot be interpreted as a statistical distribution of the physical properties, even though it correctly predicts the statistical distribution for the outcomes of all possible measurements. The question that needs to be asked is this: how is it possible that a formalism predicts all possible realities without also providing a joint reality of all physical properties? As I will explain in the following, the answer to this question can be found by considering recent experimental results on weak measurement statistics and on measurement uncertainties \cite{Lun11,Lun12,Erh12,Roz12,Kan14,Bam14}. Quantum mechanics can then be understood as a modification of the fundamental relations between physical properties \cite{Hof14}.

\section{Empirical reality and the role of measurement uncertainties}

The reason why we tend to believe in the reality of physical objects is that we can see and touch them. In physics, we extrapolate this belief to include other forms of measurement as well. However, we should keep in mind that the reality of a physical property can only be established through a process of interaction that results in an observable effect outside of the object. In quantum mechanics, this requirement is critical, since the effects of the interaction appear to limit the resolution of possible measurements. It is therefore impossible to directly access the reality of a system by simultaneously measuring a complete set of physical properties. 

Originally, it was thought that the uncertainty limits of quantum measurements provided a valid excuse for the failure to explain quantum mechanics in terms of the actual physical properties of individual systems. However, quantum mechanics does say a lot more about the relation between different physical properties than the uncertainty principle suggests. Experimentally, these relations can be explored by weak measurements, where the measurement interaction is so weak that the disturbance of the system can be neglected. The result of the weak measurement is then symmetrically defined by the combination of initial and final conditions \cite{Aha88}. Alternatively, it is possible to analyze the precise statistical structure of measurement errors. As shown by Ozawa, this results in much lower uncertainty limits if the initial information of the input state is included in the evaluation of the measurement \cite{Oza03}. Interestingly, the error statistics defined by Ozawa correspond to the statistics observed in weak measurements \cite{Hal04,Lun10,Hof12a}. The results of weak measurements are therefore consistent with a much wider range of joint measurements, as shown by a number of alternative experimental approaches \cite{Hof12b,Suz12,Hof12c,Hir13}. The statistical analysis of experimental results thus indicates that the fundamental relations between three non-commuting physical properties are accurately described by the complex conditional probabilities given by the weak values of projection operators,
\begin{equation}
\label{eq:comprob}
P(m|a,b) = \frac{\langle b \mid m \rangle \langle m \mid a \rangle}{\langle b \mid a \rangle}.
\end{equation} 
Here, the initial condition $a$ and the final condition $b$ define the actual reality of the system, while $m$ is a potential reality corresponding to an alternative measurement. In classical physics, $m$ would simply be a function of $a$ and $b$, and the reality $(a,b)$ would uniquely define the value of $m$. However, the complex conditional probability in Eq.(\ref{eq:comprob}) defines the relation between $(a,b)$ and $m$ by assigning a complex phase that describes the action of transformation from $a$ to $b$ along $m$ \cite{Hof11}. Complex conditional probabilities therefore replace and correct the classical assumption of a simultaneous reality of $a$, $b$ and $m$ with a fundamentally different concept of determinism \cite{Hof12d}. 

As shown in \cite{Hof14}, it is possible to derive all of quantum mechanics by replacing the inaccurate assumptions of classical determinism with equally deterministic complex probabilities that are governed by the law of quantum ergodicity. This new law of physics describes a fundamental relation between transformation dynamics and empirical reality, where static realities only emerge as a result of dynamical averaging. In the following, I will discuss how the law of quantum ergodicity relates to the conventional notion of states and how it fundamentally modifies the classical concept of motion. Quantum physics can then be understood as a new insight into the nature of time and motion, replacing the interpretational ambiguities of the abstract formalism with a complete set of empirically valid statements about observable reality \cite{Hof14}.

\section{Quantum ergodicity and the properties of stationary states}

Originally, the concept of states was introduced in order to describe the ``stationary'' orbits of electrons in the hydrogen atom. In the classical limit, these stationary states correspond to time-averaged trajectories, where a time interval $dt$ corresponds to a well-defined line segment $(dx,dp)$ in phase space. For a closed orbit of period $T$, the ergodic probabilities of a state of energy $E$ can be expressed in terms of the phase space distribution of position $x$ and momentum $p$,
\begin{equation}
\label{eq:class}
\rho(x,p|E) = \frac{1}{T} \; \delta(E-H(x,p)).
\end{equation}
Here, the classical Hamiltonian $H(x,p)$ provides an expression of energy as a function of position and momentum. Effectively, the classical ergodic average is based on the assumption that the joint reality of $x$ and $p$ uniquely determine a single value of energy $E$ that changes continuously if $x$ or $p$ are varied. In quantum mechanics, this interpretation of phase space points as joint realities of all physical properties breaks down. It is therefore necessary to find a different statistical expression for the ergodic averages represented by stationary quantum states. 

It is possible to approach the problem by using weak measurements of position followed by a strong measurement of momentum (or vice versa). Since the weak values of projection operators are generally complex, such measurements result in complex joint probabilities for the ergodic probabilities of a given state \cite{Lun12,Bam14,Hof12d}. In terms of the Hilbert space formalism, the quantum ergodic probabilities for a state of energy $E$ are given by
\begin{equation}
\label{eq:quant}
\rho(x,p|E) = \langle p \mid x \rangle \langle x \mid E \rangle \langle E \mid p \rangle. 
\end{equation}
Importantly, this result shows how quantum mechanics modifies the deterministic relation between energy $E$ and the phase space coordinates $(x,p)$. It is therefore possible to identify the essential principle of quantum mechanics by considering the difference between the classical determinism of $E=H(x,p)$ and the quantum determinism described by the complex valued probabilities of Eq.(\ref{eq:quant}).

Essentially, quantum mechanics replaces the joint reality of $E$, $x$ and $p$ with complex conditional probabilities $P(x|E,p)$ that relate the actual realities of $E$ and $p$ with a potential reality of $x$. Since the conditional probability of position $x$ is obtained from the noisy statistics of a weak measurement, the individual results have very little meaning and do not define an empirical reality for the systems prepared in $E$ and measured in $p$. An actual measurement of $x$ would completely randomize both energy $E$ and momentum $p$, resulting in a different ergodic average along the trajectory defined by $x$. The relation between the complex conditional probabilities $P(x|E,p)$ and the ergodic probabilities observed in a sequential measurements of $x$ and $p$ is given by the law of quantum ergodicity as introduced in \cite{Hof14},
\begin{equation}
\label{eq:qergo}
|P(x|E,p)|^2 P(p|E) = P(p|x) P(x|E).
\end{equation}
Clearly, this relation can never be satisfied by the classical ergodic probabilities of Eq.(\ref{eq:class}), since a specific value of $p$ selects only a specific set of points along the phase space trajectory of $E$, without any relation to the global ergodic averages described by $P(x|E)$. Instead, quantum ergodicity assigns a complex phase to represent the deterministic relation between $x$, $E$ and $p$. As discussed in \cite{Hof11}, this complex phase is given by the action of transformation $S(x,p,E)$ between $p$ and $E$ along trajectories of constant $x$, where the constant $\hbar$ describes the ratio between action and phase. The complex conditional probability describing the deterministic relation between $x$ and $(E,p)$ can therefore be written as  
\begin{equation}
\label{eq:cprob}
P(x|E,p) = \exp\left(i\frac{S(x,p,E)}{\hbar}\right) \sqrt{\frac{P(p|x) P(x|E)}{P(p|E)}}.
\end{equation}
The classical limit can be recovered by coarse graining the position $x$, since the gradient of the action $S(E,p,x)$ corresponds to the classical difference between the momentum of $(E,x)$ and the momentum $p$,
\begin{equation}
\label{eq:jacobi}
\frac{\partial S(x,p,E)}{\partial x} \approx f_p(x,E) - p,
\end{equation}
where $f_p$ is the classical solution of the momentum for $H(x,f_p)=E$. For coarse graining intervals of $\Delta x$, the probability $P(x|E,p)$ averages out if $(f_p-p)\Delta x \gg \hbar$. Classical determinism emerges as an approximation of quantum determinism if the product of uncertainty in position and momentum are much larger than the action-phase ratio $\hbar$. 

\section{The emergence of Hilbert space}

As shown in \cite{Hof14}, the law of quantum ergodicity can be explained in terms of conventional probability theory, without any reference to Hilbert space concepts. It is therefore misleading to present Hilbert space vectors and the associated wavefunctions as fundamental physics. Instead, the wavefunction $\psi_E(x)=\langle x \mid E \rangle$ should be explained in terms of the physics of complex conditional probabilities from which it originates. 

The misconception that state vectors are somehow fundamental to quantum mechanics can be traced to the appearance of the squared absolute value of a complex conditional probability in the law of quantum ergodicity given in Eq. (\ref{eq:qergo}). If the goal is the prediction of measurement probabilities $P(x|E)$ for the time-averaged stationary state $E$, the law of quantum ergodicity can be reformulated as
\begin{equation}
P(x|E) = \left| \sqrt{\frac{P(p|E)}{P(p|x)}} P(x|E,p) \right|^2.
\end{equation}
Significantly, this relation indicates that the ergodic probability $P(x|E)$ can be obtained without any time averaging integral from the conditional and ergodic probabilities of a single reference momentum $p$. In general, any reference momentum can be used. The conventional definition of the wavefunction emerges if the reference momentum is zero, 
\begin{equation}
\label{eq:psi}
\psi_E(x) = \sqrt{\frac{P(p=0|E)}{P(p=0|x)}} \; P(x|E,p=0).
\end{equation}
Note that this relation does not provide the probability amplitude $\psi_E(x)$ with any particular physical meaning. Instead, it indicates that the algebra of Hilbert space tends to obscure the fundamental laws of physics expressed by complex conditional probabilities. This problem is illustrated by the arbitrary but non-trivial choice of $p=0$, which is necessary for a proper physical definition of phases in the Hilbert space formalism. The fact that a different choice of reference momentum changes the phases of the wavefunction shows that the expression $\psi_E(x)$ cannot be understood properly without referring to the third property $p=p_0$ in the more fundamental complex conditional probabilities from which the Hilbert space concepts originate. 

Since the wavefunction can now be derived from a more fundamental relation between physical properties, it is also possible to understand the formal concepts of superposition and of quantum interference in terms of their physical origin. In fact, the commonly used statement that ``a system is prepared in a superposition'' is somewhat misleading: all quantum states $\mid \psi_E \rangle$ are ultimately defined by an ergodic average associated with some generating property $E$. As shown by Eq.(\ref{eq:psi}), the mathematical definition of $\psi_E$ as a superposition of different $x$ is really an expression of the deterministic relation between the property $E$, the property $x$, and a reference $p_0$. Although this expression can be used to define an unknown physical property $E$ in terms of the known phase space properties $x$ and $p=0$, this does not mean that ``$E$ is a superposition of different $x$''. Instead, the correct explanation is that ``$E$ is a property related to $x$ and $p_0$ by $P(x|E,p_0)$''. Likewise, quantum interference is not an interference of alternative realities or parallel worlds. Instead, the possibility of defining different properties $m$ in terms of complex conditional probabilities $P(m|x,p_0)$ can be used to find a direct relation between $m$ and $E$ by using a statistical chain rule,
\begin{equation}
P(m|E,p_0) = \int P(m|x,p_0) P(x|E,p_0) dx.
\end{equation}
In this integral, the physical property $x$ is used to determine the physics of $m$ and $E$. Since the relations with $x$ and $p_0$ fully determine $m$ and $E$, it is possible to derive the relation between $m$ and $E$ by integrating over the fundamental relations with $x$. One could in fact say that quantum interference is a somewhat misleading description of quantum determinism: the complex conditional probabilities should not be interpreted in terms of hidden realities, but as representations of deterministic causality equivalent to classical trajectories. In particular, this means that the deterministic relations between physical properties at different times must be formulated in terms of complex conditional probabilities to obtain the correct quantum mechanical description of motion.

\section{Time and motion}

In classical physics, the time evolution of a closed system can be expressed by deterministic functions of the initial conditions. For example, the position $x_t$ of a particle at time $t$ can be expressed as a function of position $x_0$ and momentum $p_0$ at $t=0$. The law of quantum ergodicity states that such relations are approximations obtained from the more fundamental complex conditional probabilities that express the correct relation between $x=t$ and $(x_0,p_0)$. In the case of non-relativist propagation in free space, the one-dimensional motion of a particle of mass $m$ is therefore described by
\begin{eqnarray}
\label{eq:motion1}
\lefteqn{P(x_t|x_0,p_0)=}
\nonumber \\ &&
\sqrt{-i\frac{m}{2 \pi \hbar t}} \exp\left(i \frac{m}{2 \hbar t}
\left(x_t-x_0-\frac{1}{m} p_0 t \right)^2 \right).
\end{eqnarray}
This expression replaces the mathematical description of trajectories as straight lines in space and time. Specifically, it is not possible to assign a constant velocity to the propagation, since $p_0/m$ is not equal to the ratio of distance and time $(x_t-x_0)/t$. The classical notion of velocity is an approximation that applies only when the uncertainties in position and momentum are much larger than $\hbar$. 

It is also possible to formulate the laws of motion by relating an intermediate position $x_m$ to the initial position $x_i$ and the final position $x_f$. If the time intervals $t_{im}$ and $t_{mf}$ are equal, motion in a straight line requires that $x_m=(x_i+x_f)/2$. However, quantum ergodicity describes this deterministic relation in terms of the possible transformations between $x_i$ and $x_f$ along constant $x_m$. The laws of motion are then expressed by
\begin{eqnarray}
\label{eq:motion2}
\lefteqn{P(x_m|x_i,x_f)=}
\nonumber \\ &&
\sqrt{-i\frac{m}{\pi \hbar T}} \exp\left(i \frac{m}{\hbar T}
\left(x_m-\frac{x_i+x_f}{2}\right)^2 \right),
\end{eqnarray}
where $T=t_{im}=t_{mf}$. In the absence of interactions, the laws of motion are governed by the same rules that apply to the relations between physical properties at a fixed time $t$. In general, the dynamics of a closed system are already included in the static description provided by a complete set of phase space coordinates. Importantly, the law of quantum ergodicity distinguishes between the internal time evolution of a system and the disturbance of the system by external forces. The former can be included in the description of the system, while the latter represents transformations of the system caused by interactions with the environment. This distinction is an important consequence of the role of interactions in the definition of empirical reality: the time evolution of a closed system is not accessible to observation and should not be described in terms of a sequence of measurement outcomes. Oppositely, the empirical notion of time arises from measurement sequences that necessarily disturb the motion of the object. The fundamental relations between observable properties make it difficult to bridge the gap between these two aspects of time, indicating that a more precise definition of the concept of time may be necessary. 

\section{Fictitious realities}

By replacing the classical laws of motion, complex conditional probabilities provide a fully deterministic explanation of all quantum phenomena. At the same time, complex conditional probabilities indicate that there is no joint reality of the three physical properties that they relate to each other. Instead, the relation between complex phases and the action of transformations ensures that the measurement interaction necessary for the emergence of a real effect also transforms the reality associated with past effects into the potential realities associated with future effects. 

The classical misinterpretation of reality clearly originates from the smallness of $\hbar$. The idea that reality can be described by differential geometry with time as an additional dimension is a speculation based on the assumption that dynamics is scale invariant. However, $\hbar$ defines a fundamental action scale, and the experimental results show that this scale relates the empirical reality of physical systems with their dynamics. As a consequence, it is not correct to describe the motion of a particle by a continuous sequence of points in space. Instead, the most fundamental description of motion is given by complex conditional probabilities such as the one shown in Eq.(\ref{eq:motion2}), where the classical trajectories emerge only as an approximation in the limit of low resolution. Specifically, the action-phase ratio $\hbar$ defines the threshold at which the approximate separation of elements of reality from their dynamics is possible. Below this threshold, the experimental evidence shows that the ``elements of reality'' describing the observable effects of an object cannot be separated from the interactions by which they are observed. 

Importantly, there is no scientific justification for the claim that reality should be independent of the dynamics of interaction. In the world we live in, the reality of an object can only be known by the effects of its interactions. The fact that the approximate separation of reality and dynamics works in everyday life is merely a consequence of the smallness of $\hbar$. The law of quantum ergodicity shows how this separation can be achieved by reducing the resolution of observations. Complex conditional probabilities are therefore entirely consistent with our everyday experience of objective reality, just as general relativity is consistent with the fact that we do not usually notice the small differences in the passage of time at different altitudes in the gravitational potential of earth. The idea that motion must be described by mathematical lines in space and time is as wrong as the idea that time should be the same for all observers. In order to understand quantum physics, we need to realize that the assumption that reality needs to be described by four dimensional differential geometries is neither intuitively nor logically necessary. Instead, the experimental evidence suggests that the proper relations between elements of reality are described by complex conditional probabilities that include the dynamics of the system in the form of action-phases scaled by $\hbar$. 

\section{Conclusions}

A proper explanation of quantum physics is only possible if the theory can be expressed in terms of empirically valid concepts. Based on recent experimental results, I have shown that all of quantum physics can be derived by replacing the deterministic relations of quantum physics with the complex conditional probabilities observed in weak measurements \cite{Hof14}. Quantum physics can then be understood as a more precise description of empirical reality based on the realization that the reality of an object cannot be separates from the interaction dynamics by which it is observed. As explained in \cite{Hof14}, the general form of complex conditional probabilities is expressed by the law of quantum ergodicity, which relates the complex conditional probabilities to the dynamically averaged ergodic probabilities observed in precise quantum measurements. It is then possible to reformulate quantum physics in terms of universally valid relations between physical properties, resulting in new insights into the fundamental structure of time and space. In particular, it is necessary to abandon the idea that time is merely a continuous sequence of instantaneous ``snap shot'' realities, since the emergence of realities necessarily include the dynamics of the system at the level of the action-phase ration $\hbar$. Complex conditional probabilities explain how the equations of motion need to be modified to accommodate the fundamental relation between empirical realities and interaction dynamics and show how the approximate description by continuous trajectories emerges in the limit of action uncertainties much larger than $\hbar$. 

This work was supported by JSPS KAKENHI Grant Number 24540427.

%%\vfill

\end{document}